\documentclass[
 reprint,
 superscriptaddress,          
]{revtex4-2}

\usepackage{hyperref}
\usepackage{graphicx}
\usepackage{dcolumn}
\usepackage{bm}
\usepackage{amsmath}
\usepackage{xcolor}

\usepackage{xcolor}
\usepackage{upgreek} 
\usepackage{verbatim}

\newcommand{\KJ}[1]{{\color{black}#1}}

\begin{document}


\title{An Analytical and AI-discovered Stable, Accurate, and Generalizable Subgrid-scale Closure for Geophysical Turbulence}

\author{Karan Jakhar}
\affiliation{Department of Geophysical Sciences, University of Chicago}
\affiliation{Department of Mechanical Engineering, Rice University}

\author{Yifei Guan}
\affiliation{Department of Geophysical Sciences, University of Chicago}
\affiliation{Department of Mechanical Engineering, Union College}

\author{Pedram Hassanzadeh}
\affiliation{Department of Geophysical Sciences and Committee on Computational and Applied Mathematics, University of Chicago}
\email{pedramh@uchicago.edu}

\date{\today}

\begin{abstract}
By combining AI and fluid physics, we discover a closed-form closure for 2D turbulence from small direct numerical simulation (DNS) data. Large-eddy simulation (LES) with this closure is accurate and stable, reproducing DNS statistics including those of extremes. We also show that the new closure could be derived from a 4th-order truncated Taylor expansion. Prior analytical and AI-based work only found the 2nd-order expansion, which led to unstable LES. The additional terms emerge only when inter-scale energy transfer is considered alongside standard reconstruction criterion in the sparse-equation discovery. 
\end{abstract}
\maketitle



Subgrid-scale (SGS) closures, or parameterizations, are essential for simulating real-world systems involving turbulent flows \cite{pope2000turbulent,sagaut2006large}. A prominent example is climate modeling, where closures are required for simulating the geophysical turbulence in the atmosphere and ocean, along with various other nonlinear, multi-scale processes \cite{randall2003breaking,hewitt2020resolving,fox2019challenges}. Despite over a century of research, a general framework for deriving turbulence closures from first principles has remained elusive, resulting in the widespread reliance on semi-empirical and ad-hoc models \cite{sagaut2006large, meneveau2000scale,duraisamy2019turbulence}. In fact, the shortcomings of these closures are the main source of epistemic uncertainty in climate change projections and weather forecasts, especially for extreme events \cite{palmer2014climate,slingo2011uncertainty,lai2024machine,bracco2025machine}. This has led to extensive efforts in recent decades to pursue innovative physical, mathematical, and, recently, artificial intelligence (AI) tools to develop SGS closures for geophysical flows \cite{berner2017stochastic, gentine2018could, marston2023recent, bracco2025machine}.

An ideal closure 1) should accurately capture the detailed structure of the SGS fluxes, 2) should accurately reproduce the interactions between the SGS processes and the resolved, large-scale dynamics, e.g., the inter-scale energy and enstrophy transfers, and 3) when coupled to the numerical solver of the resolved scales, e.g., in large-eddy simulation (LES), the simulated flow should have key characteristics such as energy and enstrophy spectra matching those of the direct numerical simulation (DNS). In the context of weather and climate prediction, of particular interest is also capturing the statistics of rare (extreme) events. {\it However, currently, a framework for developing closures satisfying (1)-(3) is lacking} \cite{pope2000turbulent,sagaut2006large,meneveau2000scale}. 

The ``structural modeling'' approach \cite{sagaut2006large}, e.g., based on truncated Taylor-series expansion of the SGS flux, yields closures such as the nonlinear gradient model (NGM2), which satisfies (1) with above 0.9 pattern correlation with filtered DNS \KJ{(FDNS)} \cite{leonard1975energy,clark1979evaluation, leonard1997large}. However, these closures fail (2)-(3), e.g., producing no inter-scale energy transfer in 2D turbulence and leading to unstable LES~\cite{carati1995representation, eyink2006multi, zanna2020data, jakhar2024learning}. The ``functional modeling'' approach \cite{lilly1967representation, germano1991dynamic, lilly1992proposed, leith1996stochastic} leads to eddy-viscosity closures such as Smagorinsky \cite{smagorinsky1963general} and Leith \cite{leith1996stochastic}, which produce stable LES and some aspects of (2) though they are often overly dissipative and miss backscattering (the transfer of energy from the SGS processes to the resolved scales), an important process in atmospheric and oceanic flows \cite{shutts2005kinetic, chen2006physical, jansen2014parameterizing, jansen2015energy,grooms2015numerical, khani2016backscatter, hewitt2020resolving, ross2023benchmarking, khani2023gradient, kang2023neural, perezhogin2024stable}. Excessive dissipation also degrades the representation of extreme events \cite{guan2022stable,guan2024online}. Furthermore, these closures fail (1), e.g., lower than 0.5 pattern correlation \cite{carati2001modelling,guan2022stable, moser2021statistical}.   

Recently, AI approaches for developing data-driven closures have received significant attention. Most efforts have focused on supervised learning with FDNS data, which is an approach akin to structural modeling \cite{duraisamy2019turbulence, brunton2020machine, bracco2025machine, eyring2024pushing}. While some studies have demonstrated that these data-driven closures can satisfy conditions (1)–(3), they suffer from three major limitations: (i) the need for large amounts of DNS snapshots for training, (ii) lack of interpretability, and (iii) poor generalization to out-of-distribution regimes, e.g., extrapolation to much higher Reynolds numbers \cite{maulik2018data, beck2019deep, guan2022stable, subel2022explaining,srinivasan2024turbulence}. \KJ{Other AI frameworks have also shown success, including}  self-supervised approaches, akin to functional modeling, which use only DNS statistics, \KJ{and those that directly learn the coarse-grained dynamics}~\cite{novati2021automating,bae2022scientific,kurz2023deep,wang2025coarsegrainingneuraloperators}. However, these approaches can still suffer from shortcomings (ii) and (iii).

A number of studies have pursued an alternative approach: ``equation discovery'', a class of AI techniques that aim to find a closed-form mathematical representation of the data \cite{schmidt2009distilling, udrescu2020ai, brunton2016discovering, zhang2018robust,chen2022symbolic,grundner2024data,newey2025model, cranmer2023interpretable}. The major appeals of this approach are data efficiency, interpretability, and generalizability, thus addressing the shortcomings of deep learning. In their pioneering work, Zanna and Bolton~\cite{zanna2020data} used Bayesian sparse regression to find a closed-form equation for SGS momentum stress in geophysical turbulence. They robustly discovered a closure that resembled NGM2 but found that while this closure accurately captured the detailed structure of the SGS fluxes, the LES was unstable (unless the predicted SGS flux was attenuated). Using the same methodology, \cite{jakhar2024learning} discovered closures of the same structure for SGS momentum and heat fluxes across various setups of 2D turbulence and Rayleigh-B\'enard convection and confirmed that they all exactly match the outcome of second-order truncated Taylor-series expansion; i.e., the discovered closure is NGM2. They also demonstrated that the LES instability results from NGM2's inability to capture any inter-scale energy transfer (neither diffusion nor backscattering). However, it remained unclear how a better equation discovery approach can be devised, though the need for a physics-informed loss function and a better library (e.g., through genetic programming \cite{schmidt2009distilling, ross2023benchmarking}) was speculated.   

In this Letter, we inform the criterion for ``discovery'' of a sparse representation with the fundamental understanding of turbulence physics. We robustly find, for diverse setups of 2D turbulent flows and with an expansive library, a closure that is NGM2 plus an additional term. We further show that this additional term is, in fact, the 4th-order term in the Taylor-series expansion of the SGS flux. We demonstrate that this new closure (NGM4, hereafter), accurately captures the structure of the SGS flux (pattern correlation $\approx 0.99$), accurately reproduces the inter-scale energy and enstrophy transfers, and leads to stable LES, that among other things, accurately reproduces the statistics of the rare, extreme events in diverse test cases that cover a broad range of dynamics mimicking atmospheric and oceanic turbulence.

The dimensionless continuity and momentum equations for LES of 2D turbulence are given by:
\begin{eqnarray} 
\frac{\partial \overline{u}_i}{\partial x_i} & = & 0, 
\label{eq:2d-fhit filtered uv continuity}\\
\frac{\partial \overline{u}_i}{\partial t} +\frac{\partial\overline{u}_i\overline{u}_j}{\partial x_j} & = & -\frac{\partial \overline{p}}{\partial x_i} + \frac{1}{Re}\frac{\partial^2 \overline{u}_i}{\partial x_j \partial x_j} -\frac{\partial \tau_{ij}}{\partial x_j} + \overline{\mathcal{Q}}_i. 
\label{eq:2d-fhit filtered uv momentum} \end{eqnarray}
Here, $\overline{(.)}$ represents a low-pass filtering operation, $u_i$ denotes velocity, $p$ is pressure,  $Re$ is the Reynolds number, and $\overline{\mathcal{Q}}_i$ represents a time-invariant external forcing, Rayleigh drag, and the Coriolis force. The SGS stress tensor, $\tau_{ij}=\overline{u_iu}_j-\overline{u}_i\overline{u}_j$, requires a closure model that expresses $\tau_{11}, \tau_{12}=\tau_{21},$ and $\tau_{22}, $ as a function of the resolved flow variables ($\overline{u}_i, \overline{p}$). 

We investigate four cases of 2D turbulence, generating a diverse flow dynamics that vary in dominant length scales and energy ($E=\frac{1}{2}u_iu_i$) and enstrophy ($Z=\frac{1}{2}\omega^2$) cascade regimes ($\omega$ is the vorticity); \KJ{see Appendix~A for more information about these cases and their properties.} Here, we regard direct numerical simulation (DNS) data as the ground truth \KJ{(details of the numerical solvers are in Appendix C). We} utilize FDNS data to learn closure for $\tau_{ij}$. The FDNS data are obtained by first filtering the DNS data using a Gaussian filter and then coarse-graining the results to the $4$ to $64\times$ coarser LES grid following \citet{jakhar2024learning}.
The equation discovery method used in this study is a sparsity-promoting Bayesian linear regression approach \cite{zhang2018robust, zanna2020data, mojgani2022discovery} based on the relevance vector machine (RVM) \cite{tipping2001sparse}, employed to derive closed-form closures for each element of $\tau_{ij}$ using FDNS data. This method operates on a library of basis functions, $\bm{\Phi}$, comprising linear or nonlinear combinations of relevant variables, such as velocity or its derivatives: 
\begin{eqnarray} 
\left[ \frac{\partial^{\left(q_1+q_2\right)} A}{\partial x^{q_1} \partial y^{q_2}}\right]^{p_1} \left[ \frac{\partial^{\left(q_4 + q_5\right)} B}{\partial x^{q_4} \partial y^{q_5}}\right]^{p_2} \left[ \frac{D C}{D t}\right]^{p_3},\label{eq
} 
\end{eqnarray} 
where $A, B = \overline{u}_1$ and \KJ{$\overline{u}_2$, $(x,y)=(x_1,x_2)$}, and $C = \overline{\omega}$ or an element of $\overline{S}_{ij} = \frac{1}{2}\left(\frac{\partial \overline{u}_i}{\partial x_j} + \frac{\partial \overline{u}_j}{\partial x_i} \right)$, the strain rate. Material derivative $\frac{D}{Dt} = \frac{\partial}{\partial t} + u_j\frac{\partial}{\partial x_j}$ is included to account for non-Markovian (memory) effects, motivated by Mori-Zwanzig formalism \citep{wouters2013multi,parish2017non}. 

This library is extensive, with integers $0 \le q \le 8$ and $0 \le p \le 2$, although the total derivative order is constrained to 8th (yielding a total of 930 terms in the library). To accurately compute higher-order spatial derivatives required for this study, we used arbitrary-precision methods \cite{ArbFFT}. The construction of this library is motivated by the Galilean-invariant and symmetry properties of the SGS terms; e.g., this library includes the Pope's tensors \cite{pope1975more,anstey2017deformation,gatski1993explicit,jongen1998general,lund1993parameterization,li2021data,reissmann2021application, ross2023benchmarking}.

The library must be comprehensive enough to express $\bm{{s}}$, a vectorized element of $\tau_{ij}$, as $\bm{s}^{\text{RVM}} = \bm{\Phi}\bm{c}$. The regression weights, $\bm{c}$, are optimized by minimizing the error $\text{MSE} = \lVert \mathit{S}^{\text{RVM}} - \mathit{S}^{\text{FDNS}} \rVert^2_2$, where $\mathit{S}$ consists of $n$ stacked samples of $\bm{s}$. The RVM enforces sparsity by iteratively pruning basis functions with weights exhibiting uncertainties above a predefined threshold, $\alpha$, and recalculating the MSE until all remaining functions have uncertainties below $\alpha$. A higher $\alpha$ increases model complexity but reduces the MSE.

\begin{figure*}
\includegraphics[width=1\textwidth]{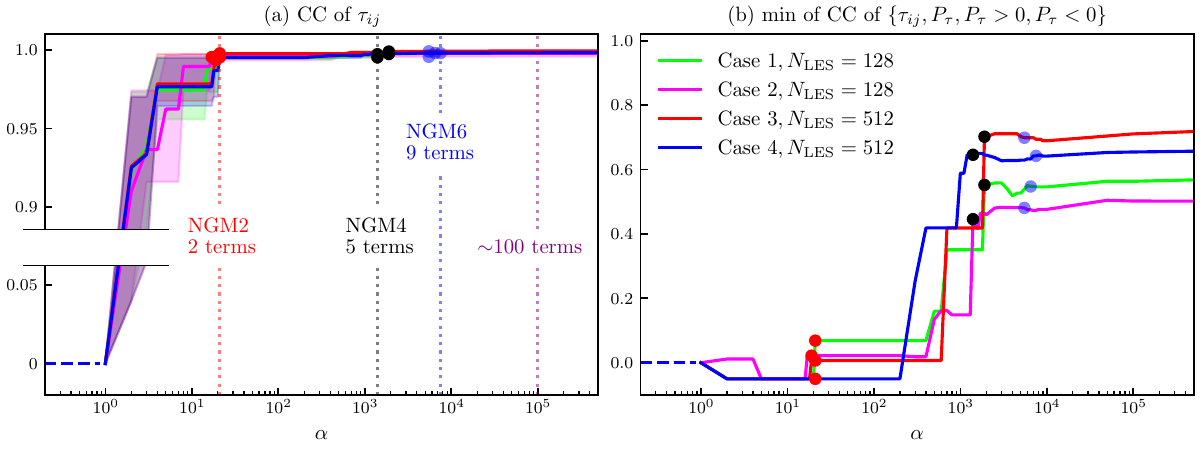}
\caption{\label{fig:CC alpha} Representative examples of the effects of increasing the sparsity-level hyperparameter, $\alpha$, on the CC in the discovered closure. (a)-(b) The CC-$\alpha$ relationship. In (a), the shading represents the max-min spread of all three elements of $\tau_{ij}$. (a) uses the common metric, CC of $\tau_{ij}$, while (b) uses a physics-informed metric that accounts for the CC of total ($P_{\tau}$), diffusion ($P_{\tau} > 0$), and backscattering ($P_{\tau} < 0$) inter-scale energy transfers as well.}
\end{figure*} 

For each of the four cases in Fig.~\ref{fig:casescontour}, we separately discover closures for three elements of the SGS stress tensor. We also examine several LES grid sizes, and the effect of varying $\alpha$, which as mentioned earlier, is a key hyper‐parameter in equation discovery. We use $n$ = 100 FDNS samples from a training set and 20 FDNS samples from an independent testing set. \KJ{We evaluate the \textit{a priori} performance using the commonly employed average correlation coefficient (CC), Eq.~\eqref{eq:CC}, on the test samples \cite{sagaut2006large,maulik2019subgrid, guan2023learning, jakhar2024learning}.}

As an illustrative example, Fig.~\ref{fig:CC alpha}(a) displays the mean CC for all elements of $\tau_{ij}$ as $\alpha$ increases. Note that this \textit{structural modeling} approach is what all past equation-discovery studies followed \cite{zanna2020data, jakhar2024learning}. For small $\alpha$  ($ < 1$), no closure is discovered (CC = 0, zero terms). As $\alpha$ increases, the number of discovered terms increases, and the CC value rises and eventually plateaus, forming an ``L-Curve.'' The elbow of this curve is commonly used to identify the $\alpha$ that balances accuracy and model size in equation discovery \cite{zhang2018robust,zanna2020data, mojgani2022discovery}. In Fig.~\ref{fig:CC alpha}(a), the CC-$\alpha$ relationship for each element of $\tau_{ij}$ converges to approximately 1 and robustly discovers the same model at the elbow. Past studies \cite{zanna2020data, jakhar2024learning} found that the elbow of this curve (red circles) corresponds to the analytically derivable NGM2 \cite{clark1979evaluation, leonard1975energy}. However, as these studies and earlier work found, LES with NGM2 leads to unstable \textit{a posteriori} (online) simulations. \citet{jakhar2024learning} suggested that this instability is due to the inability of NGM2 in capturing any inter-scale energy transfers; i.e., $P_{\tau}^{\text{NGM2}}(x_1,x_2) = 0$ (NGM2 captures neither diffusion nor backscattering). 

Here, we introduce a physics-informed discovery criterion that requires not only accurate reconstruction of $\tau_{ij}$ but also capturing the inter-scale energy transfer (both diffusion and backscattering), thus combining structural and functional modeling. Figure \ref{fig:CC alpha}(b) demonstrates that the L-curve in this new approach does \KJ{not} form an elbow around NGM2 (red circles), but rather, discovers a new closure (black circles) that has CC of $\tau_{ij}>0.99$ (Fig.~\ref{fig:CC alpha}(a)) and non-zero inter-scale energy transfer (Fig.~\ref{fig:CC alpha}(b)). We have found that the closure discovered at the new elbow is NGM2 plus the second term from the Taylor-series expansion involving the higher orders ($\mathcal{O}(\Delta^4$)), where \KJ{filter width, $\Delta$, is twice the LES grid spacing} (see Eq.~\eqref{eq:NGM}). We refer to this new closure as NGM4. As shown below, NGM4 outperforms existing physics-based closures in \textit{a priori} and \textit{a posteriori} tests.

It is noteworthy that the L-curves in Fig.~\ref{fig:CC alpha} are consistently observed for all cases with any other $N_{\text{LES}}$. Note that increasing $\alpha$ leads to another closure with more terms but negligible improvements in CC; we have found this closure to be NGM6; see Eq.~\eqref{eq:NGM} (a generalized Taylor-series expansion of $\tau_{ij}$ is provided in the supplementary information \cite{supp}).

\begin{widetext}
\begin{equation}
\tau_{ij}^{\text{NGM6}} = \underbrace{\underbrace{\frac{1}{1!}\frac{\Delta^2}{12} \left( \frac{\partial \overline{u}_i}{\partial x_k} \frac{\partial \overline{u}_j}{\partial x_k}  \right)}_{\tau_{ij}^{\text{NGM2}}} +
\frac{1}{2!}\frac{\Delta^4}{12^2}
\left(\frac{\partial^2 \overline{u}_i}{\partial x_k \partial x_l}\frac{\partial^2 \overline{u}_j}{\partial x_k \partial x_l} \right)}_{\tau_{ij}^{\text{NGM4}}} +
\frac{1}{3!}\frac{\Delta^6}{12^3}\left(\frac{\partial^3 \overline{u}_i}{\partial x_k \partial x_l \partial x_m }\frac{\partial^3 \overline{u}_j}{\partial x_k \partial x_l \partial x_m} \right) + \mathcal{O}\left(\Delta^8\right).
 \label{eq:NGM}
\end{equation}
\end{widetext}

The equation discovery with the expansive 930-term library confirms that Taylor-series expansion offers the best representation of both $\tau_{ij}$ and the inter-scale energy transfer. Note that the memory terms never emerged during the discovery.

\begin{figure*}
\includegraphics[width=1\textwidth]{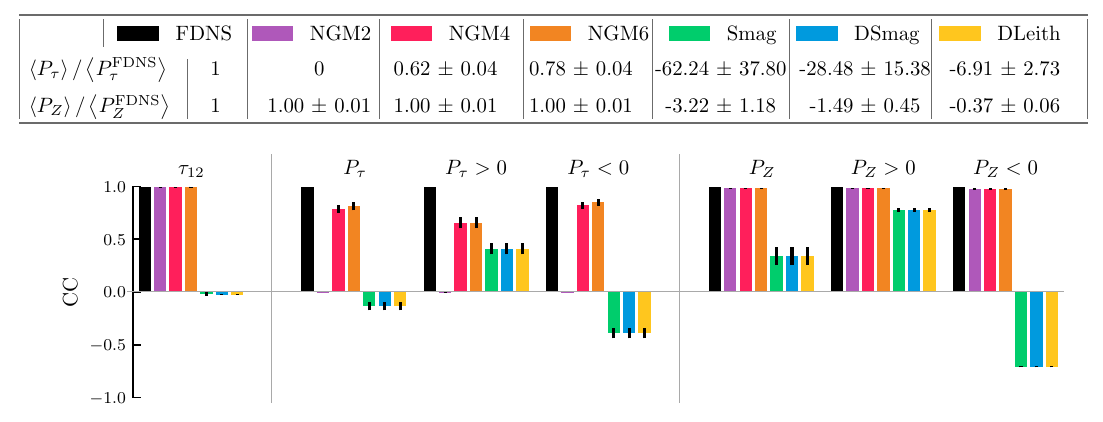}
\caption{\label{fig:apriori} Comparison of the \textit{a priori} (offline) performance of different closures. The table shows the ratio of domain-averaged $P_{\tau}$ and $P_Z$ to that of FDNS. The bar plots present the CC for $\tau_{12}$, $P_{\tau}$, and $P_Z$ \KJ{of each closure and FDNS}. All values are the mean and standard deviations calculated for \KJ{the 4 setups together} (Cases 1-2 at $N_{\text{LES}}=128$ and Cases 3-4 at $N_{\text{LES}}=512$); Table~S1 shows the values for each case.}
\end{figure*}

We first compare the \KJ{\textit{a priori} (offline)} performance of NGMs (NGM2-6) with common baseline closures, the eddy-viscosity Smagorinsky (Smag) \cite{smagorinsky1963general}, its dynamic version (DSmag) \cite{germano1991dynamic}, and the dynamic Leith (DLeith) \cite{leith1996stochastic} models (see Appendix D). Figure~\ref{fig:apriori} shows that NGM4 and NGM6 perfectly capture the structure of the \KJ{FDNS} SGS stress and inter-scale enstrophy transfers, and very well capture the inter-scale energy transfers (both diffusion and backscattering), significantly outperforming NGM2 and the physics-based closures.

An important feature of NGM4 and NGM6 is that they capture the energy backscatter that is seen in the FDNS. While the overall representations of inter-scale energy transfer of NGM4 and NGM6 are not perfect (around $62\%$ and $78\%$ of FDNS, respectively), they are substantially better than the energy transfers of NGM2 ($=0$) and the eddy-viscosity closures, which are purely and excessively diffusive (as indicated by the negative CC values). The advantages of NGM4 and 6, especially over the eddy-viscosity closures, can also be seen in the inter-scale energy and enstrophy transfers' spectra (Fig.~S1). These \textit{a priori} tests show that NGM4 and NGM6 are fairly comparable, though NGM6 has a better $P_\tau$ (see Table~S1).

\begin{figure*}
\includegraphics[width=1\textwidth]{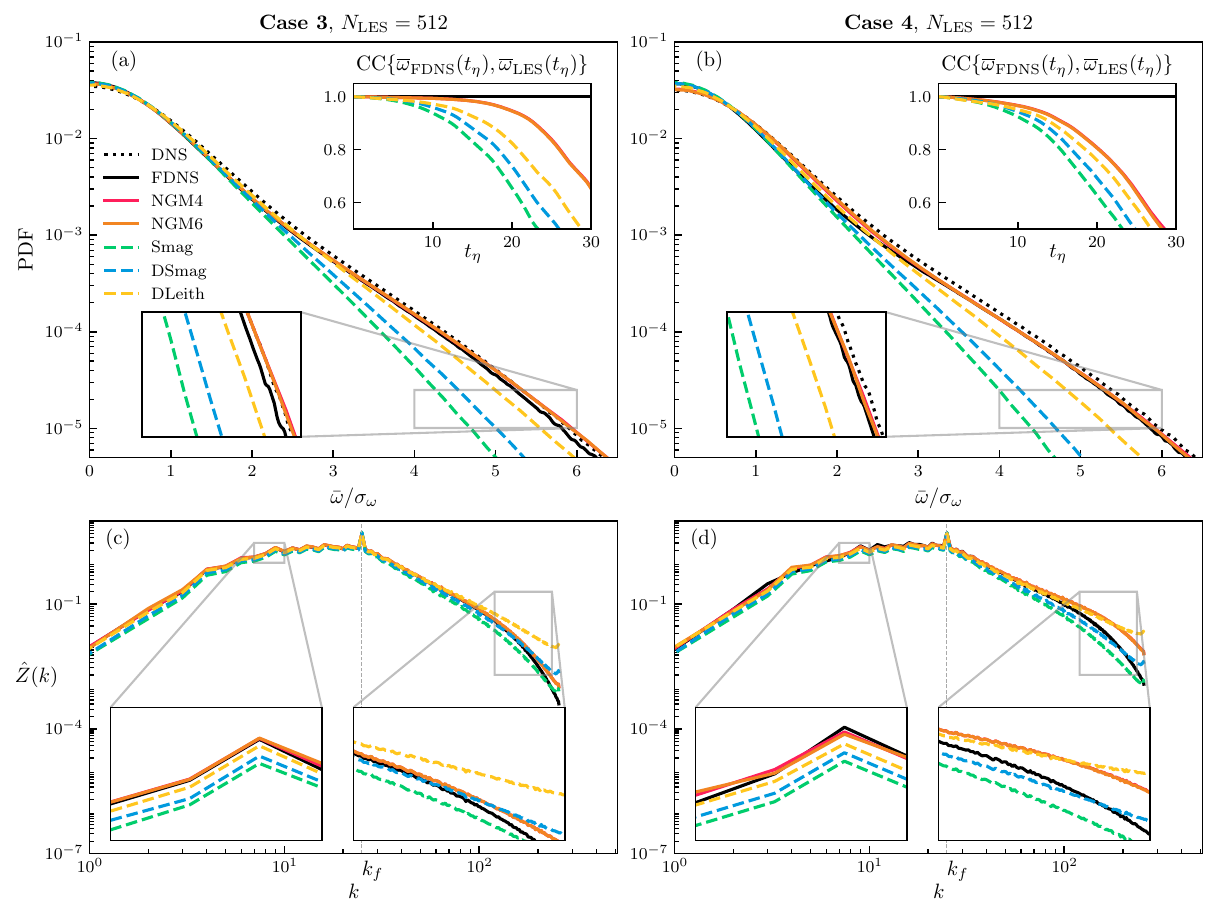}
\caption{\label{fig:aposteriori} Comparison of the \textit{a posteriori} performance of different closures for Cases 3-4 at $N_{\text{LES}}=512$. (a)-(b) \KJ{The PDF of vorticity normalized by its standard deviation}, $\overline{\omega}/\sigma_{\omega}$. The upper-right insets of (a)-(b) compares the accuracy of short-term forecasting of each LES against FDNS. The $x$-axis represents the eddy-turnover time, $t_{\eta} = 1/\sqrt{\langle\overline{\omega}^2\rangle}$. (c)-(d) Comparison of the enstrophy spectra, $\hat{Z}\left(k\right)$, of LES with different closures. A similar figure for (Cases 1-2 at $N_{\text{LES}}=128$ \KJ{and Case 4 at $N_{\text{LES}}=1024$}) is shown in the supplementary information.
}
\end{figure*}

NGM4 and NGM6's strong \textit{a priori} performance is reflected in their superior \textit{a posteriori} performance. LES \cite{py2d} with NGM4 and NGM6 remain long-term stable when the LES resolution is high enough to capture over $80\%$ of DNS enstrophy (see Fig.~S2). This is reminiscent of findings in 3D turbulence, where LES typically needs a resolution high enough to cover at least $80\%$ of DNS kinetic energy \cite{pope2000turbulent}. 

\KJ{In the atmosphere, intense cyclonic and anticyclonic vortices, which have positive and negative vorticity and are associated with low- and high-pressure systems, respectively, can drive various kinds of weather extremes \cite{vallis2017atmospheric}. Therefore, in this prototype of atmospheric turbulent circulation, the tails of the probability distribution function (PDF) of vorticity represent low-probability, rare ``extreme weather'' events}. As shown in Figs. \ref{fig:aposteriori} and S3-S4, when stable, LES with NGM4 and NGM6 capture both the bulk and tails of the PDF of vorticity. LES with eddy-viscosity closures fail to capture the extremes, especially for Cases 3 and 4, which have high-wavenumber forcings. Note that \KJ{using} NGM2 for any case and any LES resolution leads to instabilities.

In addition to better capturing extreme events, LES with NGM4 and NGM6 outperforms eddy-viscosity models in short-term forecasting of vorticity (see insets of panels (a)-(b) in Figs. \ref{fig:aposteriori} and S3-S4).  

LES with NGM4 and NGM6 also outperform LES with other closures in capturing the domain-averaged kinetic energy and enstrophy (Table~S2). Figures~\ref{fig:aposteriori}, S3 (c)-(d), \KJ{and S4 (b)} show the enstrophy spectra of FDNS and LES. The insets magnifying the tails of the spectra do not clearly show the superiority of LES with any specific closure. However, the insets magnifying the spectra at lower wavenumbers reveal that LES with eddy-viscosity closures \KJ{underestimate the enstrophy at large scales, likely due to the excessive diffusion}. The same observation can be made about the energy spectra (Fig.~S5). Note that in \textit{a posteriori} tests, LES with NGM4 and NGM6 show fairly similar performances, suggesting that in practice, using NGM4 might be enough. 

In conclusion, a physics-informed data-driven discovery identified an analytically derivable, yet previously ignored, representation of SGS closure that satisfies both \textit{structural} and \textit{functional} modeling requirements. This closure (NGM4) is ``interpretable'' as it is derivable from the Taylor-series expansion and it represents the Leonard and cross stresses (see supplementary information \cite{supp}). This closure is ``generalizable'' since its coefficients are entirely determined by the filter width ($\Delta$) and does not have any component that depends on $Re$ or any other parameters. NGM4 is accurate: it is the first closure to simultaneously excel in both \textit{a priori} and \textit{a posteriori} tests. LES with NGM4 is stable only when the numerical resolution is high enough to capture $80\%$ of DNS enstrophy; however, this is consistent with general limitations of \KJ{(explicit)} LES \cite{pope2000turbulent} \KJ{(see below for more discussions)}.

\KJ{Although this closure could have been derived analytically decades ago, AI played a critical role in its discovery here. First, the analytical expansion of the gradient model to higher-order terms had not been previously pursued. Among the main reasons are that the source of the shortcoming of NGM2 was unclear and gradient closures were generally ignored in favor of stable eddy-viscosity ones. The sparse equation discovery showed that among 930 terms, NGM4 and NGM6 best represent both the stress tensor and inter-scale energy transfer. It was only after the AI-based discovery that the connection with higher-order Taylor-series expansion became clear.}

A major difference between NGM4 and NGM6 is that the latter represents the Reynolds stress too (see supplementary information \cite{supp}). However, while \textit{a priori} tests show that NGM6 better represents the inter-scale energy transfer, \textit{a posteriori} tests do not demonstrate any advantage over NGM4 in terms of stability or accuracy. This is encouraging as implementing models with high-order derivatives in numerical solvers can be challenging; these findings suggest that NGM4 might suffice in practice, e.g., in ocean modeling. We note that NGM4 and NGM6 violate the Boussinesq hypothesis as they depend on second-order velocity gradients, highlighting the need to go beyond classical turbulence tensors to represent backscatter and anisotropic effects \cite{pope1975more}. 

\KJ{The findings of this work can also be framed within the broader context of nonlocal closures \cite{philip1968diffusion,knobloch1977diffusion,hamba2022analysis,shirian2022eddy,villermaux2025nonlocal}. The finite-order gradient closures (Eq.~\eqref{eq:NGM}) can be viewed as truncations of the nonlocal deconvolution operator \cite{sagaut2006large}. Therefore, NGM4 can be interpreted as the minimal local truncation that successfully captures the leading-order nonlocal effects required by our reconstruction and inter-scale transfer criteria. This interpretation suggests that for cases requiring stronger scale-selectivity or wider spectral fidelity, Padé-like rational approximations \cite{iliescu2003large} or explicitly nonlocal kernels can be a natural extension.}

\KJ{Furthermore, we find that NGM4, a deterministic, local closure, suffices for these cases of 2D turbulence with our chosen criteria. However, this still leaves the door open for stochastic (e.g., generative) extensions and closures with memory. In particular, applying equation-discovery methodologies to find parsimonious stochastic closures \cite{schneider2021learning,dong2025stochastic} is therefore a promising direction for future work.}

\KJ{It is also important to note that in practice, many applications, such as weather and climate modeling, use ``implicit'' LES \cite{margolin2006modeling,pressel2017numerics}. In these highly under-resolved simulations, the numerical discretization acts as an implicit closure, effectively fusing the numerical method and the SGS modeling. Our work presents a testable, resolution-aware closure that, in the explicit LES regime, is compelling for its verified inter-scale energetics and {\it a posteriori} performance. In the under-resolved implicit LES regimes, this closure can serve as a target for diagnostics of inter-scale transfers and extremes, aid interpretability studies, and potentially be combined with existing discretization schemes when an explicit representation of backscatter is desired.}

\KJ{Finally, it should be noted that} while here we only focused on the SGS stress tensor in 2D turbulence, these findings are relevant to multi-scale modeling in other dynamical systems too, as coarse-graining any quadratic nonlinearity can lead to an NGM-like representation, \KJ{e.g., as already shown for heat flux in Rayleigh-B\'enard convection~\cite{jakhar2024learning}.} This work also presents the importance of combining physical and mathematical insight to better harness the power of AI methods in accelerating scientific discovery. 

\subsection*{Acknowledgment}
\KJ{We thank two anonymous reviewers for insightful feedback.} This work is supported by NSF grant AGS-2046309 and Schmidt Sciences. Computational resources were provided by NSF ACCESS (allocation ATM170020) and NCAR’s CISL (allocation URIC0009).

\subsection*{Data availability}
The code for the numerical solver ``py2d'' is available at \cite{py2d}. The codes and data for equation discovery and analysis in this work can be found at \cite{eqsdiscovery, eqs_discovery_data}.

\subsection*{Author Contributions}
All authors participated in conceptualizing the project, designing the computational experiments, and interpreting the results. K.J. developed the codes, performed the analysis, and generated the figures. P.H. acquired the funding and supervised the research.

\bibliography{main}

\clearpage

\section*{End Matter}
\begin{figure*}[h!]
\includegraphics[width=0.95\textwidth]{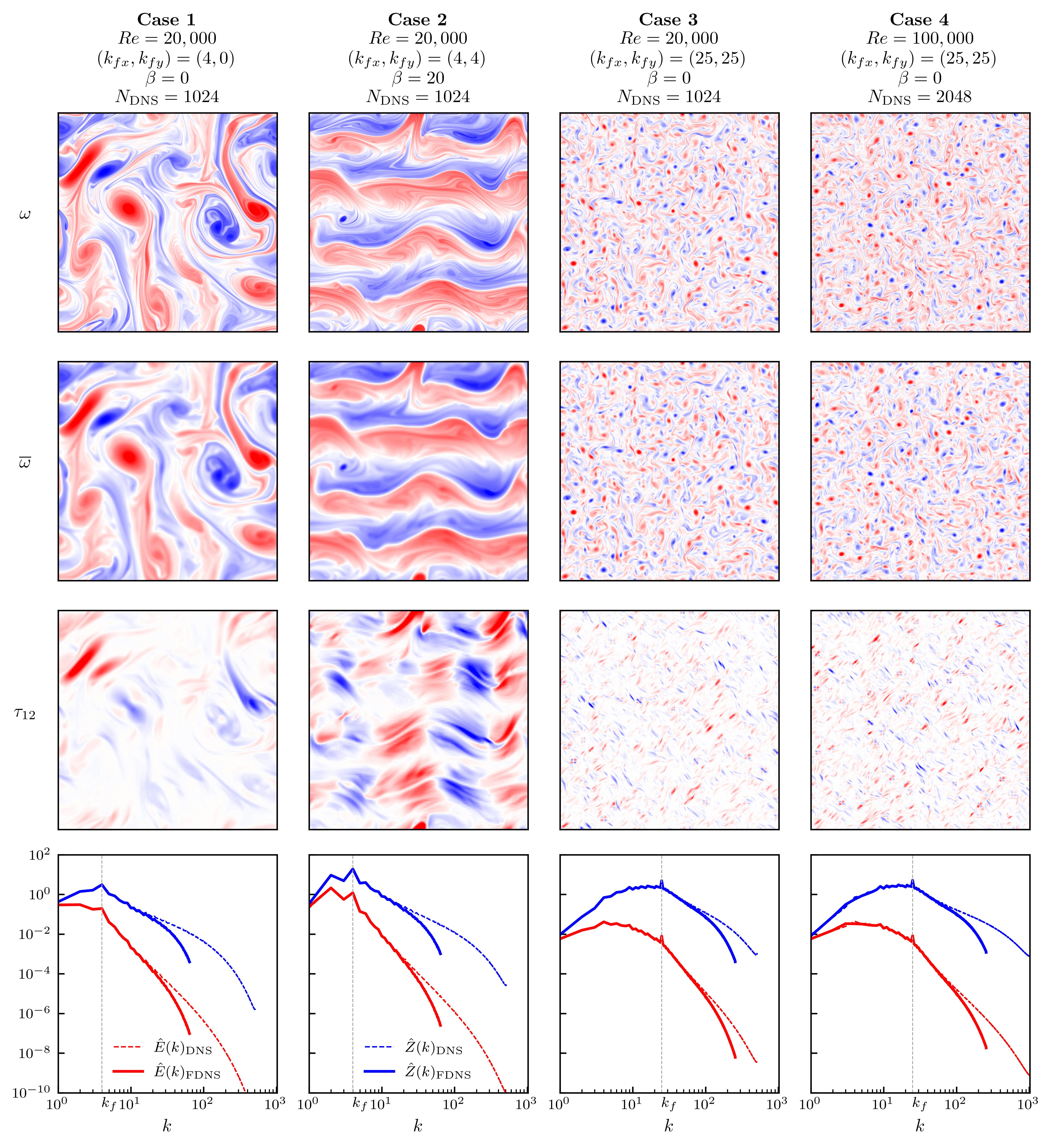}%
\caption{\label{fig:casescontour} The four cases considered in this study. These cases represent a broad range of dynamics in 2D turbulence and mimic the diversity of flow regimes (jets and vortices) in the atmosphere and ocean.  $(k_x,k_y)$ are the wavenumbers of the time-invariant forcing in the $x$ and $y$ directions, \KJ{where higher wavenumbers lead to small-scale vortices and double cascade}. $\beta$ is the gradient of the Coriolis force, \KJ{mimicking the effects of Earth's rotation and leading to the formation of jets} (see \cite{jakhar2024learning} for more details about the dynamics of these cases). $N_\mathrm{DNS}$ is the resolution of the pseudo-spectral solver (in each direction) used for DNS (\citet{py2d}). The first row presents snapshots of the DNS vorticity, $\omega$. The second row shows snapshots of the FDNS vorticity, $\overline{\omega}$. The third row displays a snapshot of an element of the SGS stress term, $\tau_{12}$ \KJ{(the subject of closure modeling)}. The last row depicts the energy (red line) and enstrophy (blue line) spectra for DNS (dashed line) and FDNS (solid line). FDNS is at $N_{\text{LES}}=128$ for Cases 1-2 and $N_{\text{LES}}=512$ for Cases 3-4.}
\end{figure*}

\textit{Appendix A}: Figure \ref{fig:casescontour} shows the cases considered in this study and their physical and numerical parameters.\\

\KJ{\textit{Appendix B}: Correlation coefficienct (CC) compares the 2D pattern of the ``true'' SGS stress, $\tau^{\text{FDNS}}$ (from FDNS), with the predicted stress, $\tau^{\text{RVM}}$, which is generated by the RVM-discovered closure based on the filtered flow variables (e.g., $\overline{u}_1$, $\overline{u}_2$) as input:
\begin{eqnarray}
    \text{CC} = \frac{\langle \left( \tau^{\text{RVM}} - \langle \tau^{\text{RVM}} \rangle \right) \left( \tau^{\text{FDNS}} - \langle \tau^{\text{FDNS}} \rangle \right)\rangle}{\sqrt{\langle \left(\tau^{\text{RVM}} - \langle \tau^{\text{RVM}}\rangle \right)^2 \rangle}\sqrt{\langle \left(\tau^{\text{FDNS}} - \langle \tau^{\text{FDNS}} \rangle \right)^2 \rangle}}, \label{eq:CC}
\end{eqnarray}
where $\langle \cdot \rangle$ is domain averaging.}\\

\KJ{\textit{Appendix C}: 
The DNS/LES numerical solver is the same as the one used in \cite{guan2022stable, jakhar2024learning}. Briefly, the Navier-Stokes equations in the vorticity-streamfunction formulation, e.g., curl of Eq.~\eqref{eq:2d-fhit filtered uv momentum} for LES,  are solved in a doubly periodic domain using a Fourier-Fourier pseudo-spectral solver. For time integration, we employ second-order Adams-Bashforth and Crank-Nicolson schemes for the advection and viscous terms, respectively.
}\\

\textit{Appendix \KJ{D}}: The baseline physics-based closures used here are Smag, DSmag, and DLeith. These eddy viscosity closures introduce dissipation and do not account for backscattering: $\tau_{ij} = -2\nu_{e}\overline{S}_{ij}$, with $\overline{S}_{ij}$ representing the filtered rate of strain and $\nu_{e}$ denoting the eddy viscosity \cite{pope2000turbulent}. $\nu_e = \left(C_s\Delta\right)^2\sqrt{2\overline{S}_{ij}\overline{S}_{ij}}$ for Smag and $\nu_{e} = \left(C_l \Delta\right)^3 |\nabla \overline{\omega}|$ for Leith. We use $C_s = 0.17$ for Smag, as proposed in a previous work \cite{lilly1967representation}; $C_s$ and $C_l$ are estimated dynamically based on the local flow structure for DSmag and DLeith, respectively. This procedure can yield $\nu_{e} \leq 0$, potentially resulting in numerical instabilities; therefore, the commonly used ``positive clipping" is applied to enforce diffusion ($\nu_e \geq 0$) \cite{maulik2019subgrid}.



\end{document}